# High-index-contrast single-mode optical waveguides fabricated on lithium niobate by photolithography assisted chemo-mechanical etching (PLACE)

Jianhao Zhang [1,2,†], Rongbo Wu [1,2†], Min Wang [3,4], Zhiwei Fang [3,4], Jintian Lin [1], Junxia Zhou [3,4], Renhong Gao [1,2], Zhe Wang [1,2], Wei Chu[3] and Ya Cheng [1,3,4,5,6,7,8*]

[1] *State Key Laboratory of High Field Laser Physics, Shanghai Institute of Optics and Fine Mechanics, Chinese Academy of Sciences, Shanghai 201800, China*
[2] *Center of Materials Science and Optoelectronics Engineering, University of Chinese Academy of Sciences, Beijing 100049, China*
[3] *State Key Laboratory of Precision Spectroscopy, East China Normal University, Shanghai 200062, China*
[4] *The Extreme Optoelectromechanics Laboratory (XXL), School of Physics and Materials Science, East China Normal University, Shanghai 200241, China*
[5] *Collaborative Innovation Center of Extreme Optics, Shanxi University, Taiyuan 030006, China*
[6] *Collaborative Innovation Center of Light Manipulations and Applications, Shandong Normal University, Jinan 250358, China.*
[7] *CAS Center for Excellence in Ultra-intense Laser Science, Shanghai 201800, China*
[8] *Shanghai Research Center for Quantum Sciences, Shanghai 201315, China*
*\* Corresponding author: ya.cheng@siom.ac.cn*

**We report fabrication of low loss single mode waveguides on lithium niobate on insulator (LNOI) cladded by a layer of SiO$_2$. Our technique, termed photolithography assisted chemo-mechanical etching (PLACE), relies on patterning of a chromium film into the mask shape by femtosecond laser micromachining and subsequent chemo-mechanical etching of the lithium niobate thin film. The high-index-contrast single mode waveguide is measured to have a propagation loss of 0.130±0.008 dB/cm. Furthermore, waveguide tapers are fabricated for boosting the coupling efficiency.**

Optical waveguides are an essential building block for photonic integrated devices. Ideally, the waveguides should enable tight mode confinement with small bend radii when aiming at the dense photonic integration applications. Meanwhile, the waveguides should also provide low propagation loss and high refractive index tunability to facilitate efficient and accurate signal transfer as well as reconfigurable signal processing. To this end, crystalline lithium niobate on insulator (LNOI) has emerged to become a material platform with promising potential [1]. Currently, low loss optical waveguides have been fabricated on LNOI using either dry etching methods such as focused ion beam milling (FIB) and lithographic technique combined with argon ion milling or a chemo-mechanical etching process (termed photolithography assisted chemo-mechanical etching (PLACE) herein) [2-13]. The LN waveguides fabricated with the dry etching can now have a remarkably low propagation loss on the order of ~0.01 dB/cm, whereas the lengths of the dry etched waveguides are limited up to a few centimeters due to the inherent low fabrication efficiencies associated with either FIB milling or electron beam lithography (EBL). The difficulty has been overcome with the development of the PLACE technique which has been used for producing low loss waveguides of a length up to ~100 cm [14-15].

Previously, the LN waveguides fabricated using the PLACE technique typically have a relatively broad top width as well as a large slant angle of sidewall with respect to the vertical direction. Due to the large mode area, the previous reported LN waveguides must be cladded with Ta$_2$O$_5$ (with a refractive index ~ 2 as compared to the refractive index ~2.2 of LN) to ensure single mode waveguiding at the telecom wavelengths [16, 17]. A propagation loss of 0.042 dB/cm have been demonstrated in the single mode LN waveguides cladded with Ta$_2$O$_5$ [17]. Meanwhile, for many applications demanding dense photonic integration, the index contrast between LN and Ta$_2$O$_5$ is too low which leads to relatively large bend radii on the level of hundreds of micrometers. Here, we report fabrication of low loss single mode LN waveguides cladded by an SiO$_2$ layer using the PLACE technique. We successfully reduce the slant angle of sidewall and the waveguide width for achieving smaller mode diameters by optimizing the chemo-mechanical etching condition. We characterize the waveguides by measuring the cross sectional profile and propagation loss. To improve the coupling efficiency at the optical interface between the fabricated waveguides and conventional lensed fibers, we fabricate waveguide tapers near the output end of the waveguides. The long-length low-loss high-index-contrast single mode waveguides fabricated on LNOI will serve as a key building block for a broad range of photonic applications.

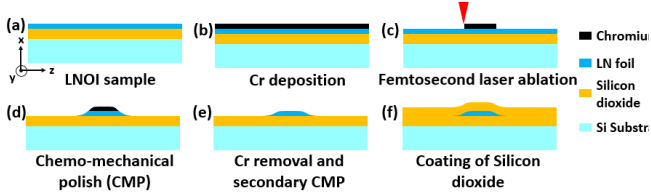

Fig. 1. Process flow of fabricating high-index-contrast single mode lithium niobate waveguides. (a) Configuration of LNOI wafer. (b) Deposition of Cr layer on top of the LN thin film. (c) Femtosecond laser micromachining of Cr film to form the stripe pattern. (d) Transferring the mask pattern from the Cr to LN by chemo-mechanical etching. (e) Removal of Cr followed by a secondary CMP. (f) Coating the LN waveguides with the $SiO_2$ cladding layer.

In our investigation, the LNOI waveguides were produced on a commercially available X-cut LNOI wafer (NANOLN, Jinan Jingzheng Electronics Co., Ltd.). The LN thin film with a thickness of 300 nm is bonded to a 4 μm-thick $SiO_2$ layer supported by a 500-μm-thick Si substrate. The configuration of the LNOI wafer is illustrated in Fig. 1(a). The process flow is illustrated in Fig. 1(b)-(f) including (1) deposition of a thin layer of chromium (Cr) with a thickness of 400 nm on the surface of the LNOI by magnetron sputtering (Fig. 1(b)); (2) patterning of Cr film using femtosecond laser ablation ((Fig. 1(c))). (3) etching of the LNOI layer by chemo-mechanical polishing (CMP) (Fig. 1(d)). (4) removal of the Cr mask left on the surface of LNOI by chemical wet etching (Fig. 1(e)). (5) deposition of silicon dioxide on the LNOI waveguide to form the cladding layer by plasma enhanced chemical vapor deposition (PECVD). The details of the femtosecond laser micromachining of Cr and the CMP etching can be found elsewhere [13,16]. However, due to the limited CMP etching ratio of the Cr mask and LN, the previous reported aspect ratio of the nanostructures fabricated with PLACE is lower than 1.5, leading to large slant angles of the waveguide sidewalls [18]. Here we applied a more stringent control on the CMP process to generate nearly vertical sidewalls of the high-index-contrast LN waveguides. Finally, the waveguide obtained with a total CMP time of 18 mins shows the narrowest top and bottom width of ~466 nm and ~625 nm, respectively, without spoiling the Cr mask as presented in its top view and cross sectional view SEM images in Fig. 2 (a) and (b), respectively. The result indicates that the PLACE fabrication technique is able to produce waveguides with geometrical dimensions comparable to the waveguides fabricated by ion dry etching. In reality, the optical waveguides used for the loss measurement in the current work has a slightly broader top width than the waveguide shown in Fig. 2 concerning the reliability and reproducibility. The high surface smoothness indicated by the SEM images is the characteristic feature of the CMP etched surface. The mode profile in the single-mode waveguide is shown by the picture captured by an infrared charge-coupled device (CCD) [InGaAs camera, HAMAMATSU Inc., Shizuoka, Japan] in Fig. 2 (c), confirming the single mode characteristic. The full width at half maximum (FWHM) of TE mode was measured to be ~0.9 μm. The measured mode profile agrees well with the simulated mode profile as shown in Fig. 2(d).

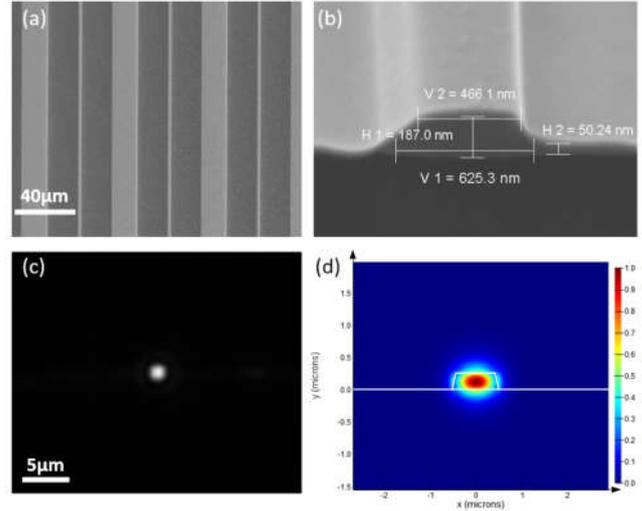

Fig. 2. (a) Top view scanning electron microscope (SEM) image of a fabricated LN waveguide (b) cross-sectional view SEM image of LN waveguide. (c) Infrared light mode field at 1550 nm in the LN waveguide. (d) Simulated TE electric field in the LN waveguide.

The SEM image of a beamsplitter based on the multi-mode interference (MMI) scheme fabricated using the same PLACE process is shown in Fig. 3(a), and the separated beams at the two output ports are shown in the inset of Fig. 3(a). A nearly 50:50 (i.e., 48.4:51.6) splitting ratio of transverse electric (TE) mode is obtained with the fabricated device which is consistent with the theoretical prediction as shown by the simulated result in Fig. 3(b). It is noteworthy that the beamsplitter based on the MMI scheme is more robust against the fabrication errors as compared with the other beamsplitting schemes [19]. The MMI-based beamsplitter can also scale up to support a large number of output ports. For these reasons, the nice agreement between the experimental and theoretical results in Fig. 3 shows the reliability and flexibility of fabrication of the MMI-based beamsplitters using the PLACE technique, as illustrated in Fig. 1, for high-density large-scale photonic integration applications.

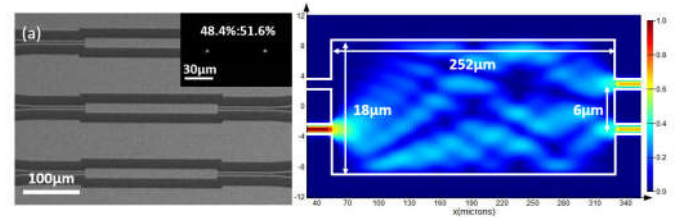

Fig. 3. (a) Top view scanning electron microscope (SEM) image of a multi-mode interference (MMI) coupler. (b) Simulated TE electric field in the MMI coupler.

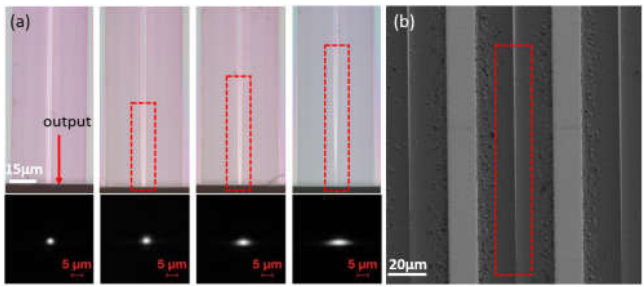

Fig. 4. (a) Optical micrographs (upper row) of the fabricated waveguide tapers of various lengths (red dashed box) and the corresponding mode profiles (lower row). (b) Top view SEM image of the rightmost waveguide taper in (a) as indicated by the red dashed box.

Another key issue in the PIC technology is the fiber-to-chip coupling, which has attracted much attention recently [20-25]. Although a high coupling efficiency of ~1.3 dB [21] has been recently demonstrated between a tapered fiber and the moderate-index-contrast LN waveguide with the $Ta_2O_5$ cladding, the same technique cannot be directly implemented on the high-index-contrast waveguides because of the much smaller mode diameters. Therefore, we fabricated waveguide tapers as shown in the optical micrograph in Fig. 4(a) for promoting the coupling efficiency between the lensed fiber (i.e., cone angle of 75) and the waveguide. This is simply achieved by selectively remove the Cr mask in the taper areas prior to the CMP etching process, thus the taper can be formed in a natural way with the CMP. The mode profiles obtained for the different taper lengths of 10 μm, 70 μm, 90 μm and 110 μm from the left to right of the upper row panels of Fig. 4(a) are shown in the lower row panels in Fig. 4(a), where the evolution of the spot from a circular spot with a diameter of ~1 μm to a highly elliptical spot with a transverse diameter of ~5 μm is observed. With the 110 μm-long waveguide taper, the expanded spot in the transverse direction has promoted the coupling efficiency from 1% to 15% as compared with the transmission rates of the waveguides with and without the taper. Higher coupling efficiencies are expected through the optimization of both the waveguide taper and the lensed fiber, which will be carefully and systematically investigated in the future.

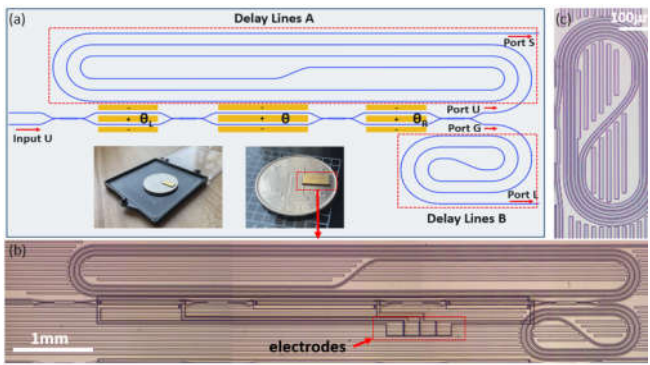

Fig. 5. (a) Schematic of the device for measuring the propagation loss. Inset: pictures of fabricated device. (b) Micrograph of the device. (c) Close up view of the coil-shaped delay lines B.

Finally, based on the photonic structures shown in Figs. 2-4, we are able to design a photonic integrated device as schematically illustrated in Fig. 5(a) for measuring the propagation loss of the high-index-contrast waveguides at high precision. The pictures of the device in the inset show its scale as compared by a one-yuan Renminbi (RMB) coin. The loss-measurement method is based on the concept of the so-called perfect Mach–Zehnder interferometer (MZI) [17, 26]. Such a perfect MZI consists of two identical electro-optic (EO) MZIs bridged with an EO phase shifter, as shown in Fig. 5(a). Due to the fabrication errors, the splitting ratio of the fabricated beamsplitters can often slightly deviate from the design. The combination of three cascaded MZIs can nicely compensate for the deviation caused by the fabrication errors, enabling production of the high-extinction-ratio beam splitters immune to fabrication imperfections. The loss measurement is easy and of high precision using such a perfect MZI. One can fabricate two waveguides of different lengths connected to the two exact 50:50 output ports of the perfect MZI and measure the different output powers at the two ports, by which the loss of the waveguide can be precisely determined. The fabricated device is shown in its optical micrograph in Fig. 5(b). The beamsplitter was fabricated on an X-cut LNOI wafer with its optical axis oriented perpendicularly to the MZI arms. After fabrication of the LNOI waveguides, Au electrodes with a thickness of 100 nm were produced by magnetron sputtering followed by space-selective patterning via femtosecond laser ablation. The half-wave voltage of the perfect MZI was recorded to be 19.92 V, and the measured extinction ratio was 29 dB. The output ports U and G of perfect MZI (indicated in Fig. 5(a)) were connected to two coil-shaped delay lines of different lengths to differentiate the propagation loss, i.e., the delay line A (upper red dashed box in Fig. 5(a)) is 40 mm longer than the delay lines B (bottom right red dashed box in Fig. 5(a)). Figure 5(c) shows the close up view of delay line B, evidencing the high density photonic integration realized in the device. Using a protocal described in detail elsewhere[17,26], the propagation loss of the single mode LN waveguide was measured to be $0.130\pm 0.008$ dB/cm by comparing the output powers from the delay lines A and B. The measured loss is higher than the single mode LN waveguide cladded with $Ta_2O_5$, most likely due to the stronger light scattering effect with the tighter confined modes in the high-index-contrast LN waveguide.

To conclude, we demonstrate fabrication of high-index-contrast single-mode LNOI waveguides with a propagation loss of $0.130\pm 0.008$ dB/cm and a mode field size of 0.9 μm. Moreover, waveguide tapers are fabricated using PLACE technique to promote the coupling efficiency from ~1% to ~15%. The result opens up the pathway for manufacturing large-scale dense PICs of high performance at low production cost.

† These authors contributed equally to this work.